\documentclass[pra,amsmath,amssymb,twocolumn,showpacs,floatfix,amsfonts]{revtex4}
\usepackage{graphicx}
\def \tr  {\rm Tr}
\def \beq {\begin{equation}}
\def \eeq {\end{equation}}

\begin{document}
\title{Quantum Zeno Effect Explains Magnetic-Sensitive Radical-Ion-Pair Reactions}
\author{I. K. Kominis}
\email{ikominis@iesl.forth.gr}

\affiliation{Department of Physics, University of Crete, Heraklion 71103, Greece}
\affiliation{Institute of Electronic Structure$\And$Laser,
Foundation for Research and Technology, Heraklion 71110, Greece}

\date{\today}

\begin{abstract}
Chemical reactions involving radical-ion pairs are ubiquitous in
biology, since not only are they at the basis of the
photosynthetic reaction chain, but are also assumed to underlie
the biochemical magnetic compass used by avian species for
navigation. Recent experiments with magnetic-sensitive
radical-ion pair reactions provided strong evidence for the
radical-ion-pair magnetoreception mechanism, verifying the
expected magnetic sensitivities and chemical product yield
changes. It is here shown that the theoretical
description of radical-ion-pair reactions used since the 70's
cannot explain the observed data, because it is based on
phenomenological equations masking quantum coherence effects. The
fundamental density matrix equation derived
here from basic quantum measurement theory considerations
naturally incorporates the quantum Zeno effect and readily
explains recent experimental observations on low- and high-magnetic-field
radical-ion-pair reactions.
\end{abstract}
\maketitle

\section{Introduction}
Recent experiments \cite{engel,fleming} on the excitonic
energy transfer taking place in photosynthetic reactions have
provided a tangible glimpse of quantum coherence effects being at
play in biological systems. The possibility of quantum physics
underpinning biological systems beyond the structural aspect has
been entertained for a long time. It has also been clear
\cite{davies,abbott} that in the decoherence-prone biological
environment, some sort of protection of the quantum coherence must
be at work before any effects related to the latter can surface.
We will here show that a familiar biophysical system, namely
radical-ion pairs and their reactions, exactly fulfills the
aforementioned requirements and exhibits effects known from
quantum physics experiments on well-isolated atomic systems and
accounted for by quantum measurement theory.

Radical-ion pairs, created by a charge transfer from a
photo-excited donor-acceptor molecular dyad, are central in the
reaction chain taking place in the photosynthetic reaction center
\cite{boxer1,boxer2}. The magnetic interactions \cite{hore,timmel}
of the two unpaired electrons in the radical-ion-pair with
external magnetic fields and internal hyperfine magnetic fields
add another layer of complexity in the charge-transfer
chain-reactions that convert the absorbed photon energy to
chemical energy vital for further biological function. Besides
photosynthesis, magnetic-sensitive recombination reactions of
radical-ion pairs are also assumed to underlie the biochemical
magnetic compass \cite{schulten1,schulten2}  used by avian and
possibly other species for navigation in the geomagnetic field
\cite{ritz,ww,ritzww,johnsen}. Furthermore, radical-ion-pairs are also
understood to participate in charge transfer reactions in DNA
helices \cite{lewis,bixon}. It is thus clear that radical-ion
pairs are found in several systems of biological significance, and
the complexity of their dynamics has been attacked from a wide
range of theoretical and experimental disciplines.

The time evolution of radical-ion-pair reactions is governed by
two distinct processes. The singlet (unpaired electron spins
anti-aligned) and triplet (unpaired electron spins aligned) states
of the radical-ion-pair and the inter-conversion between them
brought about by the magnetic interactions \cite{hore} are one piece of the
dynamics. The other is the spin-state dependent
charge-recombination of the radical-ion-pair. Singlet pairs
recombine only to singlet neutral molecules and triplet pairs
recombine only to triplet neutral molecules.

However, since the 1970's, the theoretical treatment (reviewed in
\cite{steiner}) of radical-ion-pair reactions has been based on
semi-classical/phenomenological equations involving the density
matrix describing the pair's quantum state, not unlike the early
rate-equation description of matter-light interactions.  The
latter is now known to mask important effects related to quantum
coherence, which become transparent only in light of the
full-blown quantum-mechanical description of atom-photon
interactions \cite{ct}. It is the recombination process that has
so far been treated phenomenologically in a single density matrix
equation simultaneously describing the magnetic interactions
within the radical-ion-pair. In light of quantum measurement
theory applied to radical-ion-pair recombination reactions, it is
here shown that this biological system manifests quantum
coherence, quantum jumps, the quantum Zeno effect and in principle
quantum correlations, that is, the full machinery of quantum
information science effects and concepts.

The article is organized as follows. In the following section we
will present the shortcomings of the phenomenological master
equation describing radical-ion-pair reactions so far while in
Section III we put forward the quantum-mechanically consistent
master equation. In Section IV we comment on spin relaxation, the
presence of which has in many cases conspired with the
phenomenological treatment of the dynamics to produce a
description consistent with experiments. In Section V we use the
fundamental quantum dynamic description of radical-ion-pair
reactions put forward in Section III to reproduce recent
experimental data by Hore and co-workers \cite{maeda} that clearly manifest
quantum coherence effects not accounted for by the
phenomenological theory. Finally, based on quantum measurement
theory, the derivation of the fundamental master equation
describing radical-ion-pair reactions, which is done along the
lines presented \cite{milburn,wiseman01,goan,oxtoby} by Milburn, Wiseman and co-workers on a rather
similar system, namely coupled quantum dots, is presented in
Section VI.
\section{Phenomenological Master Equation}
\begin{figure}
\includegraphics[width=8 cm]{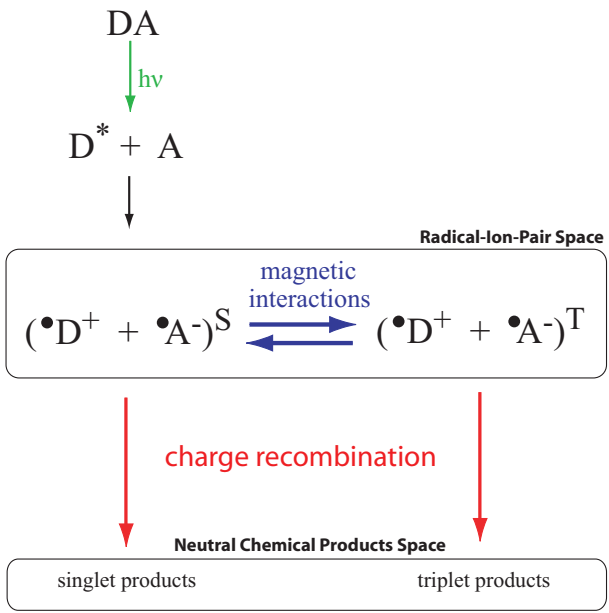}
\caption{A simple schematic of radical-ion-pair reaction dynamics.
A photon excites a donor-acceptor molecular dyad, which after a charge transfer process creates a radical-ion-pair. The magnetic interactions of the two unpaired electrons (two dots) with the external magnetic field and the molecule's magnetic
nuclei bring about a coherent singlet-triplet conversion. This
would go on forever if it were not for the charge-recombination
process that takes radical-ion pairs to neutral products. Angular
momentum conservation enforces spin selectivity of the
recombination, i.e. singlet (triplet) radical-ion pairs recombine
to singlet (triplet) chemical products.}
 \label{fig:f1}
\end{figure}
The phenomenological or semi-classical density matrix equation
that has been used so far to describe the dynamics of
radical-ion-pair reactions, schematically depicted in Fig.
\ref{fig:f1}, is \beq {{d\rho}\over {dt}}=-i[{\cal
H}_{mag},\rho]-k_{S}(\rho Q_{S}+Q_{S}\rho)-k_{T}(\rho
Q_{T}+Q_{T}\rho)\label{cl_ev} \eeq The first term describes the
unitary evolution due to the magnetic Hamiltonian ${\cal
H}_{mag}$, and the other two terms take into account the
population loss due to singlet and triplet state recombination,
respectively. The dimension of the density matrix $\rho$ is
determined by the number and the nuclear spin of the magnetic
nuclei in the radical-ion-pair. It is given by $dim=4n$, where 4
is the spin multiplicity of the two unpaired electrons and
$n=(2I_{1}+1)\times(2I_{2}+1)\times ...\times(2I_{j}+1)$ is the
spin multiplicity of the $j$ magnetic nuclei existing in the donor
and acceptor molecules. The Hermitian operators $Q_{S}$ and $Q_{T}$ are the
singlet and triplet projection operators, respectively, given by
\begin{align}
Q_{S}&=1/4-\mathbf{s}_{1}\cdot\mathbf{s}_{2} \\
Q_{T}&=3/4+\mathbf{s}_{1}\cdot\mathbf{s}_{2}
\end{align}
where $\mathbf{s}_{1}$ and $\mathbf{s}_{2}$ are the spins of the
two unpaired electrons (obviously $Q_S$ and $Q_T$ have to be
properly written as 4n-dimensional matrices operating in the
4n-dimensional Hilbert space of the particular radical-ion-pair in
consideration). The singlet and triplet projection operators are
orthogonal, $Q_{S}Q_{T}=Q_{T}Q_{S}=0$ and add up to unity,
$Q_{S}+Q_{T}=\mathbf{1}$, where $\mathbf{1}$ is the 4n-dimensional unit matrix.
Finally, being projection operators, $Q_S$ and $Q_T$ satisfy the
relations $Q_{S}^{2}=Q_{S}$ and $Q_{T}^{2}=Q_{T}$, respectively.

It is usually considered that the radical-ion-pair starts out in
the singlet state, so that we can write for the initial density
matrix $\rho(t=0)=Q_{S}/n$, where $n$ is the nuclear spin
multiplicity. We let $S(t)={\rm Tr}\{\rho(t)Q_{S}\}$  and
$T(t)={\rm Tr}\{\rho(t)Q_{T}\}$ denote the time-dependent
probability to find the radical-ion-pair in the singlet and
triplet state. Due to the structure of \eqref{cl_ev}, the trace of
the density matrix decays exponentially to zero, since from
\eqref{cl_ev} it follows that \beq {{d{\rm Tr}\{\rho\}}\over
{dt}}=-2k_{S}S-2k_{T}T\label{cl_trace} \eeq This is the main
problem of the phenomenological equation \eqref{cl_ev}, i.e. in
order to describe population loss from the radical-ion-pair space
into the neutral chemical products space due to charge
recombination, the normalization of the density matrix is forced
to an exponential decay. {\it All coherences and populations are
consequently also forced "by hand" to follow an exponential decay,
thus eliminating the actual presence of quantum coherence
effects}.

What has not been realized so far, however, it that the tunneling taking place in the charge-recombination
process is fundamentally a selective quantum measurement continuously interrogating the radical-ion-pair's spin state.
If treated as such, the recombination process reveals well-known effects from the realm of quantum measurements, such
as the quantum Zeno effect \cite{misra,pascazio}, long-lived quantum coherences and quantum jumps \cite{plenio}.

Quantum Zeno effects appear in several physical systems, some of
which are very similar to radical-ion-pairs, like the ortho-para
conversion in molecular spin isomers \cite{ortho}, ultra-cold atom
tunneling through optical potentials \cite{raizen}, or the
suppression of transverse spin-relaxation due to spin-exchange
collisions in dense alkali-metal vapors
\cite{happer_tang,kominis_pla}. In the latter case, atomic
spin-exchange collisions, of the form
$\mathbf{s}_{1}\cdot\mathbf{s}_{2}$, where $\mathbf{s}_{1}$ and
$\mathbf{s}_{2}$ are the electron spins of the two colliding
atoms, probe the atomic spin state. When the collision rate
(measurement rate) exceeds the intrinsic frequency scale of the
system, which is the Larmor frequency of spin precession in the
applied magnetic field, the effective decay rate of the spin
coherence is suppressed, a phenomenon that has led to the
development of new ultra-sensitive atomic magnetometers
\cite{romalis_prl,nature_magn}. The radical-ion-pair tunneling into the neutral
state is essentially a scattering process \cite{carminati,kubala},
not unlike atomic collisions, that performs a measurement of the
pair's spin state, since tunneling to singlet (triplet) products can only proceed if the two unpaired electrons are
in the singlet (triplet) spin state. Quantum Zeno effects have been
extensively analyzed in the literature
\cite{pascazio_review,shimizu}, both with respect to pertaining
physical systems, as well as the general conditions leading to the
quantum Zeno effect or its inverse, the anti-Zeno effect
\cite{kurizki}.
\section{Quantum Measurement Theory Description of Radical-Ion-Pair Reactions}
Using standard quantum measurement theory \cite{wiseman} it
readily follows (as shown in Section VI in detail) that the fundamental density
matrix equation describing radical-ion-pair reactions and the associated magnetic-field effects is
\begin{align}
{{d\rho}\over {dt}}=-i[{\cal H}_{mag},\rho]&-k_{S}(\rho Q_{S}+Q_{S}\rho-2Q_{S}\rho Q_{S})\nonumber\\&-k_{T}(\rho Q_{T}+Q_{T}\rho-2Q_{T}\rho Q_{T})\label{q_ev_1}
\end{align}
which is the same as the \eqref{cl_ev}, apart from the terms
$2Q_{S}\rho Q_{S}$  and $2Q_{T}\rho Q_{T}$. It is these terms that
are responsible for the quantum effects that become important in
the parameter regime where the recombination rates are larger
than the magnetic interactions frequency scale. Since
$Q_{S}+Q_{T}=1$ , \eqref{q_ev_1} can be simplified to \beq
{{d\rho}\over {dt}}=-i[{\cal H}_{mag},\rho]-k(\rho
Q_{S}+Q_{S}\rho-2Q_{S}\rho Q_{S})\label{q_ev} \eeq where
$k=k_{S}+k_{T}$. The physical interpretation of \eqref{q_ev} is
that both singlet and triplet recombination channels essentially
"measure" the same observable, $Q_{S}$, with a total measurement
rate $k_{S}+k_{T}$. In other words, the singlet (triplet)
recombination channel is interrogating the radical-ion-pair
whether it is in the singlet (triplet) state. But being in the
triplet is not being in the singlet ($Q_{S}+Q_{T}=1$), thus both
channels measure the same observable. This physical interpretation
of the actual dynamics is completely lacking from the
phenomenological equation \eqref{cl_ev}. The density matrix
equation \eqref{q_ev} has the property that the normalization of the density
matrix does not change, i.e. ${\rm Tr}\{\rho\}=S(t)+T(t)=1$ at all
times. It is here noted that as early as 1976, Haberkorn
\cite{hab} arrived at \eqref{q_ev} based on semi-quantitative
arguments, but did not further consider it, exactly because it
does not seem to describe population loss from the
radical-ion-pair space into the reaction products space due to
recombination, since ${\rm Tr}\{\rho\}=1$ at all times. All works
henceforth have used the phenomenological equation \eqref{cl_ev}.

To take recombination into account, we need another stochastic
equation involving quantum jumps \cite{wiseman} out of the
radical-ion-pair space into the neutral chemical products space. These are given by the probability of
the singlet and triplet channel recombination,
\begin{align}
p_{S}&=2k_{S}\langle Q_{S}\rangle dt\nonumber\\
p_{T}&=2k_{T}\langle Q_{T}\rangle dt\label{qjump}
\end{align}
The physical interpretation of the above equations is the
following: $p_{S}$ ($p_{T}$) is the probability that the
radical-ion-pair recombines into the singlet (triplet) channel in
the time interval between $t$ and $t+dt$. Both $p_{S}$ and $p_{T}$
are obviously positive-definite numbers (the expectation values of $Q_S$ and $Q_T$ range between 0 and 1). In other words, Haberkorn
obtained the correct density matrix evolution equation, missing,
however, the quantum jump
equations \eqref{qjump} and their physical interpretation, as the conceptual underpinnings of
quantum measurement theory and open quantum systems were lacking
at the time. In Section V on the explanation of recent experimental data we will explicitly explain how the
quantum dynamic evolution described by \eqref{q_ev} and \eqref{qjump} is actually simulated.

{\it Thus the fundamental quantum dynamical evolution of
radical-ion-pair recombination reactions is given by \eqref{q_ev}
together with \eqref{qjump}}. To summarize the fundamental
difference of the quantum description based on \eqref{q_ev} and
\eqref{qjump} versus the phenomenological \eqref{cl_ev} we note
that \eqref{cl_ev} is such that it continuously removes population
from the singlet-triplet subspace into the chemical product space,
and does so independently for each recombination channel. In
reality however, the actual evolution of the radical-ion-pair's
spin state is given by \eqref{q_ev} (both channels affect the spin
state evolution) and the molecules recombine in random times and
in jump-like fashion according to \eqref{qjump}. Thus even if the
radical-ion-pair starts out in the singlet state, a high triplet
recombination rate will significantly affect the spin state of the
molecule, well before it has time to acquire a triplet character
through the magnetic interactions. This physical statement is not
embodied in the phenomenological equation \eqref{cl_ev}.
\begin{center}
\begin{figure*}
\includegraphics[width=18 cm]{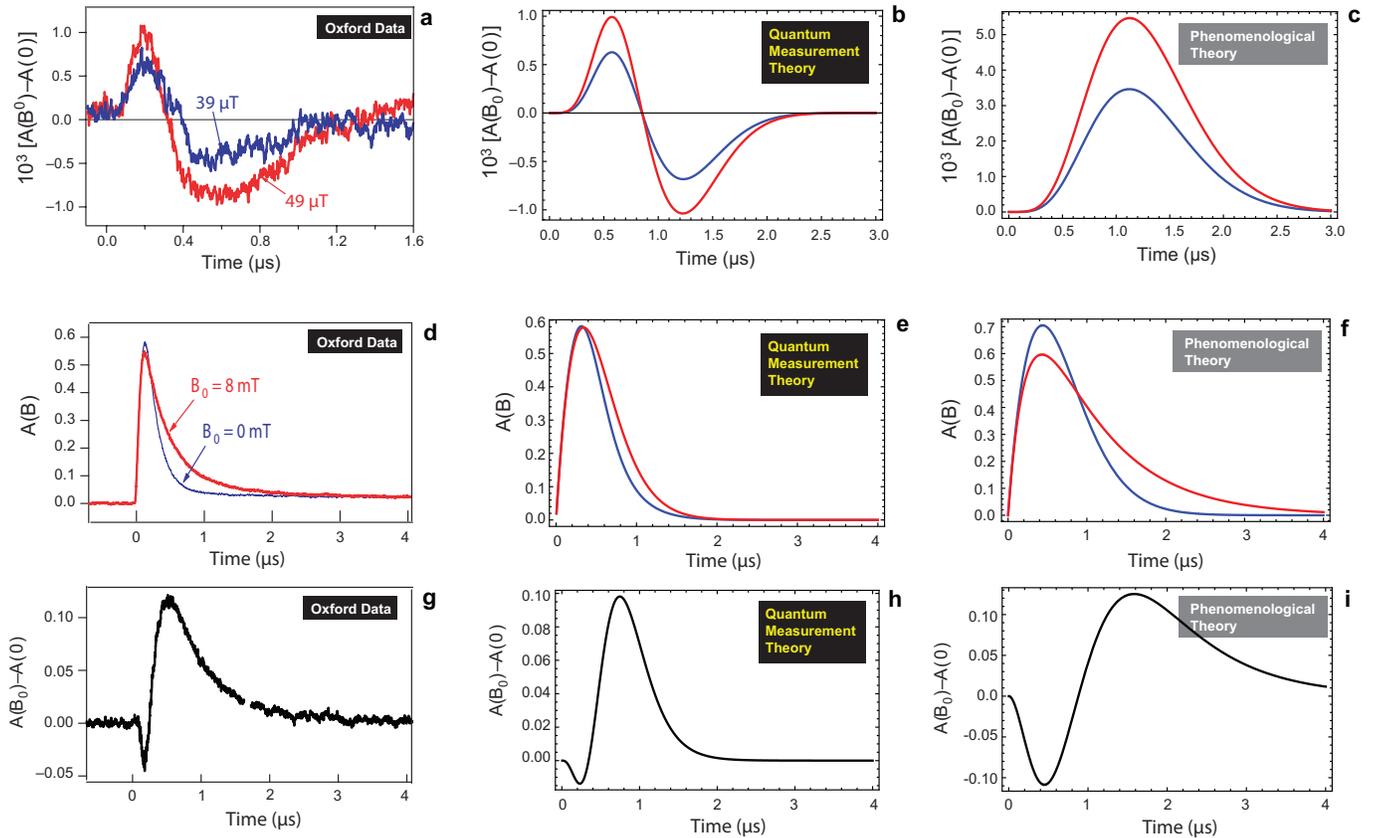}
\caption{Magnetic Field Effect, Experimental Data and Theoretical Calculation.
(a) Measured \cite{maeda} magnetic field effect for two different magnetic fields $B=49~{\rm \mu T}$
(red curves) and $B=39~{\rm \mu T}$ (blue curves). (b) Reproduction of data from quantum measurement theory based on
\eqref{q_ev} and \eqref{qjump} with a simple one-nuclear-spin (spin-1/2)
radical-ion-pair model with a diagonal hyperfine tensor given by
$a_{x}=8~\mu {\rm s}^{-1}$, $a_{y}=2~\mu {\rm s}^{-1}$ and
$a_{z}=0$. The recombination rates used are $k_{S}=0.05~{\rm \mu
s^{-1}}$ and $k_{T}=3.5~{\rm \mu s^{-1}}$. (c) Reproduction of data from the phenomenological theory \eqref{cl_ev} with the
same parameters as in (b). (d) Measured \cite{maeda} transient absorption traces at zero and high magnetic field and (g) corresponding
magnetic field effect. (e) and (h) Reproduction of data from quantum measurement theory based on
\eqref{q_ev} and \eqref{qjump} with the same parameters as in (b). (f) and (i) Reproduction of data from the phenomenological theory \eqref{cl_ev} with the same parameters as in (b). For the reproduction of the high field data
we had to consider the creation dynamics of radical-ion
pairs (mentioned in Section IV) with $k_{cr} =4~{\rm \mu s}^{-1}$
and a high field relaxation rate $k_{B,sr} =1.0~{\rm \mu s}^{-1}$. As explained in Section IV, due to the spin
relaxation present at high fields, the qualitative disagreement
between data and phenomenological theory is not that pronounced as
at low fields.}
 \label{fig:data}
\end{figure*}
\end{center}
\section{Spin Relaxation}
There are two points that need to be commented on before
proceeding to present numerical simulations and the explanation of recent experimental data. The creation of
radical-ion pairs does not happen instantaneously, but at a given
rate $k_{cr}$. This can be modeled by adding a source term to the
density matrix equations, $k_{cr}e^{-k_{cr}r}$ , the time-integral
of which is equal to 1. This is not a fundamental point and will
henceforth be neglected, i.e. we will assume that the rate
$k_{cr}$ is much larger than all other rates of the problem,
unless noted otherwise. Second and more important is the effect of
spin-relaxation. This can be modeled by a term
$-k_{sr}(\rho-\rho_{0})$ , where $k_{sr}$ is the relaxation rate
and $\rho_{0}=\mathbf{1}/4n$  is the fully mixed density matrix, i.e. a
diagonal $4n$-dimensional matrix with diagonal elements equal to $1/4n$. In the
phenomenological equation \eqref{cl_ev} the analogous
spin-relaxation term should read $-k_{sr}\rho$, since by design
equation \eqref{cl_ev} has $\rho=0$ as the infinite-time solution,
hence the relaxation term should work towards the same steady
state. We will "turn on" spin relaxation by setting $k_{sr}\neq
0$, to show that in the presence of spin relaxation the results
following from the quantum-mechanical description based on
\eqref{q_ev} and \eqref{qjump} naturally blend in with the ones
derived from \eqref{cl_ev}. This is the main reason why so far the
phenomenological density matrix equation \eqref{cl_ev} has provided
a more or less consistent theoretical description of experimental
observations. Obviously, the charge recombination process itself
induces spin relaxation, and this is what is described by
\eqref{q_ev}. By the rate $k_{sr}$ we describe all other spin
relaxation effects, for example spin-lattice relaxation. It is
also noted that diffusion or other effects add another layer of
complexity to the dynamics described by \eqref{q_ev} and
\eqref{qjump} and will not be considered here. In other words,
equations \eqref{q_ev} and \eqref{qjump} are the fundamental
dynamical equations describing magnetic-sensitive radical-ion-pair
reactions in the idealized case of negligible diffusion. This is
not unrealistic, since at low temperatures (as for example in the
experiment reported in Ref. \cite{maeda}) the effects of diffusion
and more importantly, spin relaxation, are suppressed.

\section{Explanation of Experimental Data}
We will now show that the quantum-mechanical analysis of the radical-ion-pair reactions
put forward in Section III seamlessly explains recent experimental data, while the phenomenological
theory is unable to do so. In the experiment recently reported \cite{maeda} by Hore and co-workers, a
carotenoid-porphyrin-fullerene (CPF) triad was used to study the
magnetic field effect on the radical-pair's recombination
dynamics. We have performed numerical simulations using the
simplest physically realizable radical-ion-pair model, that containing just one
spin-1/2 magnetic nucleus. In this case the density matrix is
8-dimensional and the magnetic Hamiltonian reads
\beq
{\cal H}_{mag}=\omega (s_{1z}+s_{2z})+\mathbf{I}\cdot\mathbf{A}\cdot\mathbf{s}_{1}\label{eq:Hmag}
\eeq
where we considered the external magnetic field
to be along the $z$-axis, $\mathbf{B}=B\mathbf{\hat{z}}$, and the
electron Larmor frequency parameter (the real Larmor frequency is
two times higher) is $\omega=\gamma_{e}B$, with $\gamma=1.4~{\rm
MHz/G}$. The hyperfine coupling tensor $\mathbf{A}$ couples the
single nuclear spin existing in either the donor or the acceptor
molecule with the corresponding unpaired electron. Anisotropic
hyperfine interactions embodied in the hyperfine tensor are
responsible for the directional sensitivity of the reaction, i.e.
the sensitivity of the reaction yields on the angle between the
external magnetic field and the local coordinate frame defined by
the hyperfine tensor. In the experiment reported in \cite{maeda}
the angular dependence of the reaction has also been measured, and
can also be reproduced theoretically from the quantum dynamic
description of the reaction, however, we will present this topic in
a different manuscript.

By direct numerical integration of either \eqref{cl_ev} or
\eqref{q_ev} and \eqref{qjump} we calculate the population of the
triad molecules in the singlet-triplet subspace at time $t$. This
population is what is measured in the transient absorption
measurement. The "magnetic-field effect" is the difference
$(S(t)+T(t))_{B}-(S(t)+T(t))_{B=0}$ , i.e. the difference of the
measured transient absorption at a magnetic field $B$ from that at
zero field. For the sake of completeness we note that while the
phenomenological predictions are obtained by a direct integration
of \eqref{cl_ev}, the quantum dynamic predictions based on
\eqref{q_ev} and the quantum jump equation \eqref{qjump} are
obtained as follows: we start with $N_0$ radical-ion pairs, and in
each time step the number of existing radical-ion-pairs $N(t)$
evolves according to \beq N(t)=N(t-dt)(1-p_{S}-p_{T}) \eeq i.e.
the number of radical-ion-pairs disappearing to the chemical
product space (either singlet or triplet) between time $t-dt$ and
time $t$ is $N(t-dt)$ times the probability to jump to a
recombined state (either singlet or triplet) during the time
interval $dt$. This probability is just $p_{S}+p_{T}$, where
$p_{S}$ and $p_{T}$ are given by \eqref{qjump}. Along the reaction, the
density matrix $\rho$ describing the spin state of the radical-ion-pair is evolved according to \eqref{q_ev}.

We take the radical-ion-pair to be in the singlet state initially,
i.e. $\rho(t=0)=Q_{S}/2$ (in general for a singlet initial state it is $\rho(t=0)=Q_{S}/n$, where $n$ is the nuclear spin multiplicity, which in
this case of one spin-1/2 nucleus is $n=2$). In Fig. 2a we show the measured magnetic
field effect at low magnetic fields (earth's field), while in Fig.
\ref{fig:data}b we fully reproduce the data using the
quantum dynamics description of \eqref{q_ev} and
\eqref{qjump}, whereas Fig.\ref{fig:data}c shows that the
phenomenological equation \eqref{cl_ev} is not capable of
reproducing the observed magnetic field effect with the same
parameters, and the same is true for any choice of the system's
parameters.

It is seen that the quantum dynamic description based on
\eqref{q_ev} and \eqref{qjump} agrees with the data both
qualitatively and quantitatively (the magnitude of the effect),
whereas the opposite is true for the phenomenological predictions
shown in Fig. \ref{fig:data}c. It is noted that the time-scale of
the measured magnetic field effect is missed by about a factor of
2, since in reality the molecule's dynamics are determined by tens
of nuclear spins, whereas here we have considered just one nuclear
spin. It is also important to note the following: in Fig. \ref{fig:data}b we
reproduce the data using a small value for the singlet-channel
recombination rate and a much larger value for the triplet-channel
recombination rate, while the molecule starts out in the singlet
state. This is exactly the parameter regime where the quantum Zeno
effect is actually manifested, i.e. as mentioned before, the high
triplet recombination rate has a tangible effect on the spin
dynamics, as it appears in the total measurement rate
$k=k_{S}+k_{T}$. This is a regime that the phenomenological theory fails to account for even at the qualitative level.
\begin{center}
\begin{figure*}
\includegraphics[width=14 cm]{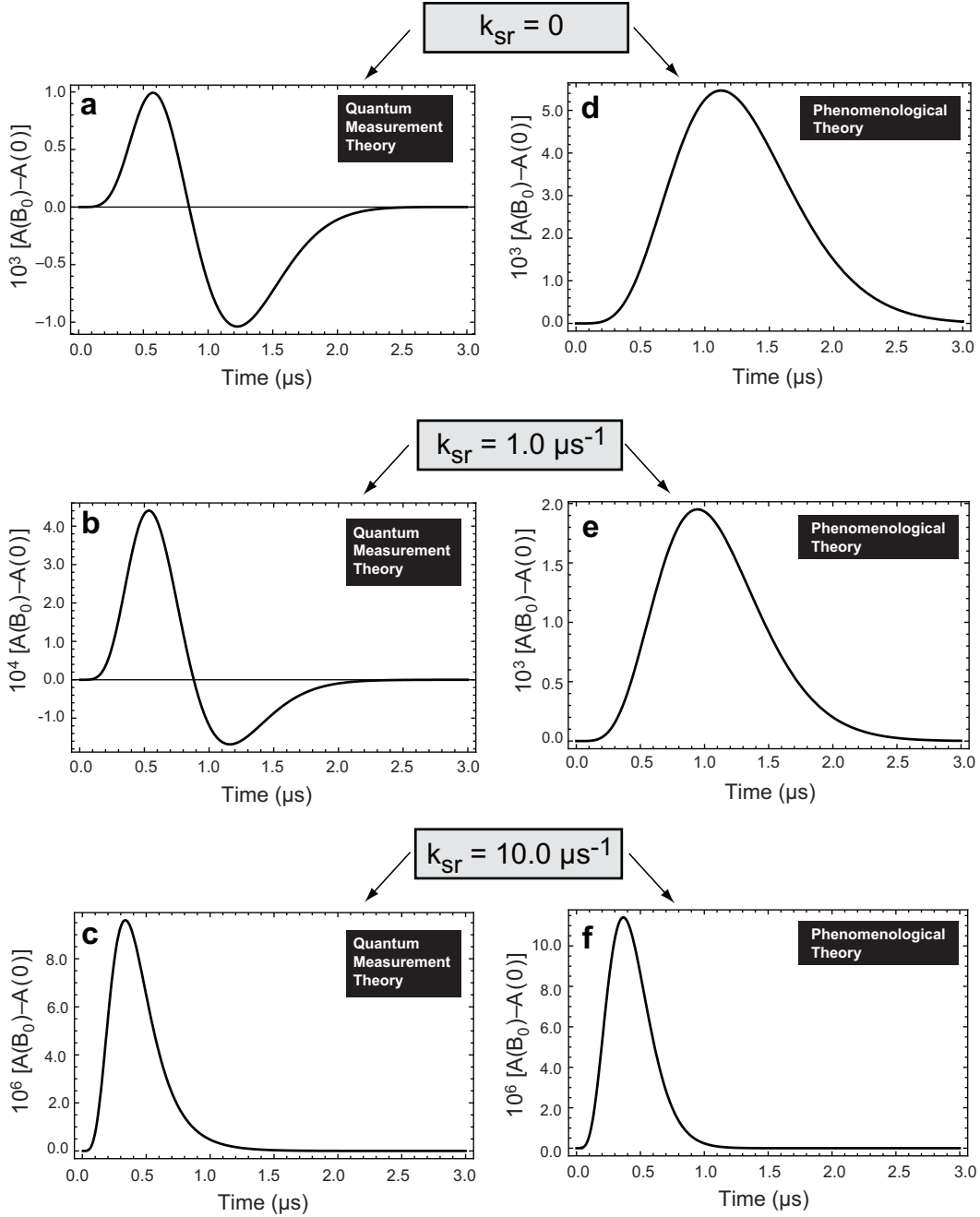}
\caption{Magnetic Field Effect and Spin Relaxation. The magnetic field effect is simulated
at a magnetic field $B=49~{\rm \mu T}$. Traces
a,b,c have been produced with the quantum dynamics given by
\eqref{q_ev} and \eqref{qjump}, while traces d,e,f have been
produced by the phenomenological equation \eqref{cl_ev}. The same
calculations have been done for three different values of the
relaxation rate $k_{sr}=0,1,10~{\rm \mu s}^{-1}$. Plot (a) is the
same as plot (b) of Fig. \ref{fig:data}, i.e. the one reproducing
the low-field data of \cite{maeda}. It is seen that by increasing
the spin relaxation rate, the two theories produce qualitatively
and quantitatively similar results, i.e. for increasing $k_{sr}$
the bi-phasic response of the magnetic-field-effect is suppressed
and the magnitude of the effect is roughly the same in both
theories. It is thus clear that the presence of spin relaxation
has in many cases masked the insufficiency of the phenomenological
description and made it appear consistent with experiments. The
other way around, it is clear that the low temperatures used in
the recent experiment by Hore and co-workers \cite{maeda} has
suppressed spin-relaxation effects to the extent that the true
quantum nature of radical-ion-pair reactions could surface.}
\label{fig:rel}
\end{figure*}
\end{center}
We will now demonstrate that if we take into account a possible
presence of spin-relaxation, the magnetic field effect is
suppressed both in the phenomenological description and in the
quantum-mechanical, and both descriptions qualitatively merge into
each other. The presence of spin-relaxation effects is one reason
why the phenomenological reaction dynamics of equation
\eqref{cl_ev} have been relatively successful until now. This is expected
since in general relaxation effects tend to equalize level
populations and damp quantum coherences, adversely affecting the
precision of any spectroscopic measurement, such as the
determination of the applied magnetic field.  We will further show
that the bi-phasic response observed in the data (Fig.
\ref{fig:data}a) and reproduced theoretically (Fig.
\ref{fig:data}b) fades away when we "turn on" spin relaxation, and
resembles the response produced by the phenomenological equation
\eqref{cl_ev}. The reason that the experiment by Maeda et al
\cite{maeda} manifests effects not accountable by the density
matrix equation used until now is the suppressed presence of
spin-relaxation effects at the low temperatures at which the
experiment was carried out. Indeed, in Fig. \ref{fig:rel} we turn
on spin-relaxation and it is
evident that the magnetic field effect derived from \eqref{q_ev}
and \eqref{qjump} resembles that derived from \eqref{cl_ev} as the
spin relaxation rate $k_{sr}$ is increased.

Finally, also the high-field data of \cite{maeda} can be explained
with the same magnetic Hamiltonian used to reproduce the low-field
data shown in Fig. \ref{fig:data}a. To this end we take into
account the creation of radical-ion pairs, which does not take
place instantaneously, but at a given rate $k_{cr}$, modelled by
the term $dN/dt=N_{0}k_{cr}e^{-k_{cr}t}$, where $N_0$ is the total
number of molecules.  In Fig. \ref{fig:data}d we show the
experimental results for the transient absorption at a magnetic
field of $B$=0 and $B$=8 mT, with the quantum measurement
calculation shown in Fig. \ref{fig:data}e and the phenomenological
prediction in Fig. \ref{fig:data}f. Similarly, Figures
\ref{fig:data}g,h,i show the data and theoretical predictions for
the magnetic field effect, i.e. the difference of the traces shown
in Figures \ref{fig:data}d,e and f. The initial negative response
is due to non-zero spin-relaxation at high magnetic fields, which
we model as described previously and which has been mentioned to
exist in \cite{maeda}. It is interesting to note the fact that the
data reproduction by quantum measurement theory is capable of
picking up this small spin relaxation ($k_{sr}=1~\mu{\rm s}^{-1}$)
existing at high magnetic fields. Taking into account the
simplicity of the magnetic Hamiltonian used for the calculations,
the agreement between experimental data and the predictions of
quantum measurement theory is excellent, as both the absolute
value of the transient absorption as well as the strength of the magnetic
field effect and the time scale of the phenomenon are perfectly
reproduced. It is also interesting to note that at high magnetic fields the
disagreement between data and the predictions of the phenomenological theory is not that
pronounced as at low fields. This is exactly due to the small spin relaxation present at high fields.
\section{Derivation of Master Equation}
In Figure 4 we depict in more detail the generic model for the radical-ion-pair creation and recombination dynamics shown in Fig. 1.
A donor-acceptor molecule DA is photo-excited (D$^{*}$A) and a subsequent
charge-transfer creates the radical-ion-pair (D$^{+}$A$^{-}$). The singlet and
triplet states of the radical-ion-pair ($^{S}$D$^{+}$A$^{-}$,
$^{T}$D$^{+}$A$^{-}$) are split by internal magnetic interactions
of the radical-ion-pair's two unpaired electrons with external magnetic fields
and internal hyperfine couplings. The radical-ion-pair is initially created in
the singlet state, which is not an eigenstate of the magnetic
Hamiltonian, and therefore a singlet-triplet (S-T) coherent mixing
commences. A tunneling event into a vibrationally-excited state of the
neutral recombined molecule DA, which quickly decays into the
ground state, completes the photon-initiated radical-ion-pair reaction.

As is well known \cite{jortner,kobori}, electron transfer in radical-ion-pair
recombination reactions is fundamentally a quantum-mechanical
tunneling process. We will show that this process
constitutes a continuous quantum measurement of the pair's spin
state. Like every quantum measurement, this one is no exception to
the rule that measurements performed on a quantum system lead to
decoherence \cite{braginsky}. However, under appropriate
conditions involving the measurement rate and the intrinsic (magnetic)
frequency scale of the radical-ion-pair, the quantum Zeno effect \cite{misra,itano}
appears and leads to physical consequences completely against the intuition that has
developed over the years based on the phenomenological description of \eqref{cl_ev}.

In the following, we are going to capitalize on the remarkably
strong analogy between radical-ion-pairs and yet another physical
system, namely two coupled quantum dots
\cite{gurvitz,milburn,wiseman01,goan,oxtoby} being continuously
interrogated by a point contact. This analogy is schematically depicted in Fig. \ref{fig:dots}. We are going to identify the
analogous physical observables of the two systems and then derive
the corresponding evolution equation describing radical-ion-pair reactions. The
electron hopping between the two dots is the analog of the S-T
coherent mixing taking place in the radical-ion-pair, whereas the measurement
performed by the point contact corresponds to the
spin-state-dependent electron tunneling into an adjacent excited state
of the neutral molecule.
\begin{figure}
\includegraphics[width=6 cm]{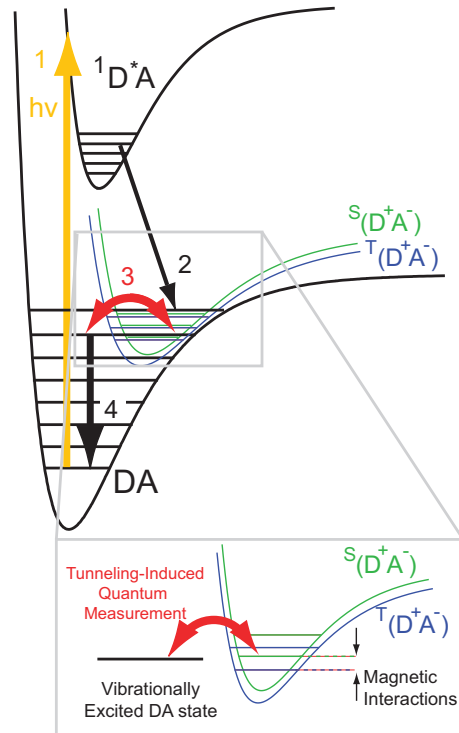}
\caption{Generic level structure and recombination dynamics in a
radical-ion-pair, taking place in four steps: 1, photoexcitation,
2, radical-ion-pair creation, 3, tunneling-induced quantum
measurement of the pair's spin state and 4, final decay to the
ground state. Only the singlet reservoir of vibrationally excited
states (of the neutral molecule DA) is shown for simplicity.
Another such reservoir exists for the triplet chemical products.}
\label{fig:potentials}
\end{figure}
Radical-ion pairs in the singlet (triplet) state will tunnel to a nearby empty reservoir of singlet (triplet) vibrationally excited states, the
annihilation and creation operators of which are denoted by $a$ ($b$) and $a^{\dagger}$ ($b^{\dagger}$), respectively.
The corresponding Hamiltonian is
\beq
{\cal H}_{res}=\sum_{a}{\omega_{a}a^{\dagger}a}+\sum_{b}{\omega_{b}b^{\dagger}b}
\eeq
As explained before, the radical-ion-pair consists of two unpaired electrons, one at the site of the donor (D) and one at the site of the acceptor (A).
We denote by $c$ and $c^{\dagger}$ the fermion annihilation and creation operators of the electron at the acceptor site.
Obviously, if $\langle c^{\dagger}c\rangle=1$ we have a radical-ion-pair, if however, $\langle c^{\dagger}c\rangle=0$, we have
a neutral recombined molecule.
\begin{figure}
\includegraphics[width=8 cm]{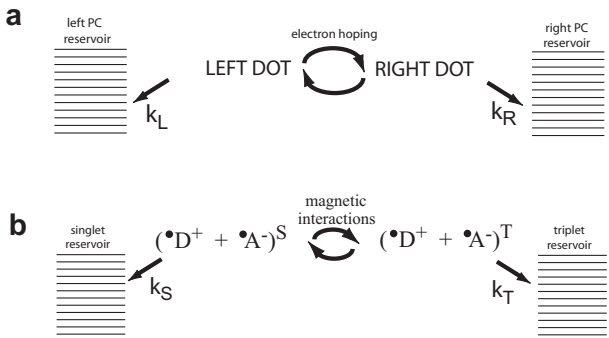}
\caption{Analogy between coupled quantum dots and radical-ion-pair
reactions. (a) Schematic of coupled quantum dots interrogated by a
double point contact (PC). (b) Schematic of radical-ion-pair with
the singlet and triplet recombination reservoirs.}
 \label{fig:dots}
\end{figure}
The Hamiltonian of the radical-ion-pair thus reads \beq {\cal
H}_{\rm rip}=c^{\dagger}c(\epsilon+{\cal H}_{mag}) \eeq where
${\cal H}_{mag}$ is the magnetic Hamiltonian operating in the spin
space of the radical-ion-pair (for a single nuclear spin it is
given in (\ref{eq:Hmag}) and it is readily generalized for any
number of nuclear spins by adding more hyperfine interaction
terms) and $\epsilon$ the energy of the radical-ion-pair with
respect to the ground state neutral molecule. Lastly, the
tunneling Hamiltonian reads \beq {\cal
H}_{T}=\sum_{a}{T_{a}ca^{\dagger}Q_{S}}+\sum_{b}{T_{b}cb^{\dagger}Q_{T}}+{\rm
c.c.} \label{tunH}\eeq The interpretation of the first term is that if the
radical-ion-pair is in the spin singlet state it can charge recombine,
reducing the radical-ion-pair occupation number to zero and
creating an occupied vibrationally excited state of the singlet reservoir.
Similarly for the triplet state in the second term. In summary,
the total Hilbert space consists of the singlet and triplet reservoirs, the
occupation of the radical-ion-pair's acceptor electron and the spin space of the
radical-ion-pair. For example, in the case of singlet-channel recombination,
the initial and final quantum states coupled by ${\cal H}_{T}$
would look like $|\Psi_{i}\rangle=|\{0\},1\rangle|\chi\rangle$ and
$|\Psi_{f}\rangle=|\{n\},0\rangle|\chi_{s}\rangle$, where the
first set of quantum numbers $\{n\}$ is the excitation of the
vibrational reservoir, the second is the electron occupation
number of the acceptor site, and $|\chi\rangle$ is the spin state
of the radical-ion-pair (two electrons plus nuclei), while
$|\chi_{s}\rangle$ is the electron spin singlet state of the
neutral donor electrons with the nuclear spin state being
the same in $|\chi\rangle$ and $|\chi_{s}\rangle$. The tunneling
amplitudes are \cite{jortner} $T_{i}=V_{i}f_{i}$ ($i=a,b$), where $V_{i}$
is the electronic matrix element and $f_{i}$ the vibrational
overlap between the nuclear wavefunctions of the reservoir states
$|i\rangle$ and the radical-ion-pair.

A very similar set of Hamiltonians has already been treated by Milburn, Wiseman and
co-workers \cite{milburn,wiseman01,goan} for the case of quantum
dots, and we will here proceed along the same lines to derive the
master equation for radical-ion pairs. We write the total
Hamiltonian as ${\cal H}={\cal H}_{0}+{\cal V}$, where \beq {\cal
H}_{0}=c^{\dagger}c\epsilon+{\cal H}_{res} \eeq and the
perturbation ${\cal V}$ is \beq {\cal V}=c^{\dagger}c{\cal
H}_{mag} + {\cal H}_{T} \eeq and move to the interaction picture
defined by ${\cal H}_{0}$, in which the perturbation is ${\cal V}_{I}=e^{i{\cal H}_{0}t}Ve^{-i{\cal H}_{0}t}$ and it follows that
\begin{align}
{\cal V}_{I}(t)&=c^{\dagger}c{\cal
H}_{mag}\nonumber\\&+\sum_{a}{T_{a}ca^{\dagger}Q_{S}e^{i(\omega_{a}-\epsilon)t}}+{\rm c.c.}\nonumber\\
&+\sum_{b}{T_{b}cb^{\dagger}Q_{T}e^{i(\omega_{b}-\epsilon)t}}+{\rm c.c.}\label{VI}
\end{align}
The proof of \eqref{VI} is given in the Appendix. We will now apply time-dependent perturbation
theory up to $2^{\rm nd}$ order. The density matrix of the total system (spin state of radical-ion-pair + electron + reservoir states) is denoted by
$W$. The density matrix of interest describing the radical-ion-pair spin state will be $\rho$, and will be calculated by tracing out
the other degrees of freedom from $W$. During the time of coherent singlet-triplet mixing, i.e. as long as the radical-ion-pair
has not recombined, the reservoir states are empty, which means that the density matrix of the singlet and triplet reservoirs is
$\rho_{res}=\sum_{a}|0\rangle_{a}{}_{a}\langle 0|\otimes\sum_{b}|0\rangle_{b}{}_{b}\langle 0|$. For the same reason, the density matrix for the occupation
state of the acceptor electron is $\rho_{c}=|1\rangle_{c}{}_{c}\langle 1|$ and the total density matrix is then given by
$W(t)=\rho(t)\otimes\rho_{res}\otimes\rho_{c}$. Obviously $\rho(t)=\tr_{a,b,c}\{W(t)\}$.

Since the tunneling part of the perturbation ${\cal V}_{I}$ is linear in the reservoir operators
$a$, $a^{\dagger}$, $b$ and $b^{\dagger}$, it easily follows that in the $1^{\rm st}$-order of the perturbation expansion ($\hbar=1$)
\begin{align}
W(t+dt)&=W(t)\nonumber\\&-idt[{\cal V}_{I}(t),W(t)]\nonumber\\&
-dt\int_{0}^{t}{dt_{1}[{\cal V}_{I}(t),[{\cal V}_{I}(t_{1}),W(t)]]}\label{pert}
\end{align}
the terms in $V_{I}$ containing the tunneling amplitudes $T_{a}$ and $T_{b}$ do not contribute, because e.g. ${\rm Tr}_{a,b}\{\rho_{res}a\}=0$.
Since $\tr_{c}\{c^{\dagger}c{\cal H}_{mag}\rho_{c}\}=\langle c^{\dagger}c\rangle {\cal H}_{mag}={\cal H}_{mag}$, it follows that the
$1^{\rm st}$-order term in \eqref{pert} results in
\beq
\left({{d\rho}\over {dt}}\right )_{1^{\rm st}-{\rm order}}=-i[{\cal H}_{mag},\rho]
\eeq
We will now proceed to calculate the integral appearing in the $2^{\rm nd}$-order term of \eqref{pert}. In this term we will neglect ${\cal H}_{mag}$ with respect to ${\cal H}_{T}$, under the assumption that the measurement dynamics embodied in the tunneling Hamiltonian \eqref{tunH} dominates
the magnetic interactions, as is actually the case in the regime where the quantum Zeno effect is manifested. This approximation is validated by the fact that the resulting master equation has the ability to reproduce experimental data. We will show the calculation just for the $a$-reservoir states (the singlet reservoir) in \eqref{VI}, as the term involving the triplet reservoir ($b$-states) is dealt with in exactly the same way. In the integral appearing in \eqref{pert}, containing $V_{I}$ to $2^{\rm nd}$-order, terms involving $T_{a}$ from $V_{I}(t)$ and $T_{a}$ from $V_{I}(t_{1})$ contain the operator $c^{2}$, which is zero. The only terms that
survive are the ones containing $T_{a}$ from the $V_{I}$(t)-term and $T_{a}^{*}$ from the complex conjugate part of the $V_{I}(t_{1})$-term. These terms lead (after taking into account the fact that cross-terms involving different reservoir states $a$ and $a'$ as well as cross terms involving both reservoirs do not contribute) to an expression of the form
\begin{align}
&-dt\sum_{a}|T_{a}|^{2}\int_{0}^{t}{dt_{1}e^{i(\omega_{a}-\epsilon)(t-t_{1})}[ca^{\dagger}Q_{S},[c^{\dagger} aQ_{S},W]]}\nonumber\\&+{\rm c.c.}
\label{2nd}\end{align}
Taking into account the fact that for the empty reservoir states it holds $\langle a^{\dagger}a\rangle=0$ and $\langle aa^{\dagger}\rangle=1$, while
for the acceptor electron it is $\langle c^{\dagger}c\rangle=1$ and $\langle cc^{\dagger}\rangle=0$, it follows that when tracing out the $a$ and
$c$ degrees of freedom from the double commutator in \eqref{2nd}, we are left with the operator $\rho Q_{S}-Q_{S}\rho Q_{S}$. The complex conjugate
(c.c.) term will then offer $Q_{S}\rho-Q_{S}\rho Q_{S}$, thus in total we obtain the Lindblad form $\rho Q_{S}+Q_{S}\rho-2Q_{S}\rho Q_{S}$.
We finally consider the resonance condition $\omega_{a}\approx\epsilon$ and make the usual approximation
$\int_{0}^{t}{dt_{1}e^{i(\omega_{a}-\epsilon)(t-t_{1})}}\approx2\pi\delta(\omega_{a}-\epsilon)$, leading to the desired result, which after doing the same calculation for the triplet reservoir states $b$, reads
\beq {{d\rho}\over {dt}}=-i[{\cal
H}_{mag},\rho]-k_{S}{\cal D}[Q_{S}]\rho -k_{T}{\cal
D}[Q_{T}]\rho\eeq where the super-operator ${\cal D}[B]$ acts on
the density matrix $\rho$ according to \beq {\cal
D}[B]\rho=B^{\dagger}B\rho+\rho B^{\dagger}B-2B\rho B^{\dagger}
\eeq We have thus derived \eqref{q_ev}, with the recombination rates given by (we have re-introduced $\hbar$ for the consistency of
units)
\begin{align}
k_{S}&=(2\pi/\hbar^{2})\sum_{a}{|T_{a}|^{2}\delta(\epsilon-\omega_{a})}\\
k_{T}&=(2\pi/\hbar^{2})\sum_{b}{|T_{b}|^{2}\delta(\epsilon-\omega_{b})}\label{eq:recomb}
\end{align}
The resulting master equation \eqref{q_ev} is indeed of the Lindblad form
that follows from standard quantum measurement theory
\cite{wiseman,braginsky,steck}, when the measured observable is
$Q_{S}$ and the measurement rate is $k=k_{S}+k_{T}$, leading to
the physical interpretation that the spin-dependent recombination
process is a continuous quantum measurement of the radical ion
pair's spin state.
\section{Conclusions}
In summary, we have shown that there exists a biological system, radical-ion pairs and their reactions, in which the full machinery of quantum measurement theory can be fruitfully applied, leading to an understanding of quantum effects visible in the laboratory but masked by the phenomenological description of radical-ion-pair reactions employed until now. A fundamental quantum phenomenon, the quantum Zeno effect,
familiar from carefully prepared atomic physics experiments, is seen to participate in the complex realm of biologically significant chemical reactions, further opening up the way to the emerging field of quantum biology, i.e. the exploration of the manifestations of quantum phenomena and the application of quantum information concepts in biological systems.
\begin{acknowledgments}
I would like to acknowledge Prof. D. Anglos for several
useful and stimulating discussions.
\end{acknowledgments}
\appendix
\section*{Appendix}
\subsection{Proof of equation \eqref{VI}}
We will show that $e^{i\epsilon c^{\dagger}ct}ce^{-i\epsilon c^{\dagger}ct}=ce^{-i\epsilon t}$. Since $c$ is a fermion annihilation operator, it
is $c^{2}=0$, thus $e^{i\epsilon c^{\dagger}ct}c=c$. Furthermore, since $(c^{\dagger}c)^{n}=c^{\dagger}c$ for any integer power $n$, it follows
that $e^{-i\epsilon c^{\dagger}c t}=e^{-i\epsilon t}c^{\dagger}c + cc^{\dagger}$, leading to the desired result. In a similar fashion it can be
shown that $e^{i{\cal H}_{res}t}a^{\dagger}e^{-i{\cal H}_{res}t}=a^{\dagger}e^{i\omega_{a}t}$ and similarly for the $b$-degrees of freedom, thus
follows equation \eqref{VI}.


\begin{thebibliography}{10}

\bibitem{engel}
G. S. Engel {\it et al.}, Nature {\bf 446}, 782 (2007).

\bibitem{fleming}
H. Lee, Y. C. Cheng and G. R. Fleming, Science {\bf 316}, 1462 (2007).

\bibitem{davies}
P. C. W. Davies, Biosystems {\bf 78}, 69 (2004).

\bibitem{abbott}
D. Abbott {\it et al.}, Fluctuations and Noise Letters {\bf 8}, C5
(2008).

\bibitem{boxer1}
S. G. Boxer, Biochim. Biophys. Acta {\bf 726}, 265 (1983).

\bibitem{boxer2}
S. G. Boxer, C. E. D. Chidsey and M. G. Roelofs, Ann. Rev. Phys. Chem. {\bf 34}, 389 (1983).

\bibitem{hore}
C. R. Timmel, U. Till, B. Brocklehurst, K. A. McLaughlan and P. J.
Hore, Molec. Phys. {\bf 95}, 71 (1998);C. R. Timmel, F. Cintolesi,
B. Brocklehurst and P. J. Hore, Chem. Phys. Lett. {\bf 334}, 387
(2001).

\bibitem{timmel}
C. R. Timmel and K. B. Henbest, Phil. Trans. R. Soc. Lond. A {\bf
362}, 2573 (2004).

\bibitem{schulten1}
K. Schulten, C. E. Swenberg and A. Weller, Z. Phys. Chem. {\bf 111}, 1 (1978).

\bibitem{schulten2}
K. Schulten, Adv. Solid State Phys. {\bf 22}, 61 (1982).

\bibitem{ritz}
T. Ritz, S. Adem and K. Schulten, Biophys. J. {\bf 78}, 707 (2000).

\bibitem{ww}
W. Wiltschko and R. Wiltscko, J. Comp. Physiol. A {\bf 191}, 675 (2005).

\bibitem{ritzww}
T. Ritz, P. Thalau, J. B. Phillips, R. Wiltschko R and W.
Wiltschko, Nature {\bf 429}, 177 (2004).

\bibitem{johnsen}
S. Johnsen and K. J. Lohmann, Physics Today {\bf 61}, 29 (2008).

\bibitem{lewis}
F. D. Lewis {\it et al.}, Science {\bf 277}, 673 (1997).

\bibitem{bixon}
M. Bixon {\it et al.}, Proc. Natl. Acad. Sci. USA {\bf 96}, 11713 (1999).

\bibitem{steiner}
U. E. Steiner and T. Ulrich, Chem. Rev. {\bf 89}, 51 (1989).

\bibitem{ct}
C. Cohen-Tannoudji, J. Dupont-Roc and G. Grynberg, {\it Atom-photon interactions:basic processes and applications}, John Wiley \& Sons, 1998.

\bibitem{maeda}
K. Maeda {\it et al.}, Nature {\bf 453}, 387 (2008).

\bibitem{milburn}
H. B. Sun and G. J. Milburn, Phys. Rev. B {\bf 59}, 10748 (1999).

\bibitem{wiseman01}
H. M. Wiseman {\it et al.}, Phys. Rev. B {\bf 63}, 235308 (2001).

\bibitem{goan}
H. S. Goan, G. J. Milburn, H. M. Wiseman and H. B. Sun, Phys. Rev. B {\bf 63}, 125326 (2001).

\bibitem{oxtoby}
N. P. Oxtoby, {\it A Quantum Trajectory Approach to Realistic Measurement of Solid-State Quantum Systems}, Ph.D. Thesis, Griffith University, 2006.

\bibitem{misra}
B. Misra and E. C. G. Sudarshan, J. Math. Phys. {\bf 18}, 756 (1977).

\bibitem{pascazio}
P. Facchi and S. Pascazio, Fortschr. Phys. {\bf 49}, (2001).

\bibitem{plenio}
M. B. Plenio and P. L. Knight, Rev. Mod. Phys. {\bf 70}, 101 (1998).

\bibitem{ortho}
B. Nagels, L. J. F. Hermans and P. L. Chapovsky, Phys. Rev. Lett. {\bf 79}, 3097 (1997).

\bibitem{raizen}
S. R. Wilkinson {\it et al.}, Nature {\bf 387}, 575 (1997).

\bibitem{happer_tang}
W. Happer and H. Tang, Phys. Rev. Lett. {\bf 31}, 273 (1973).

\bibitem{kominis_pla}
I. K. Kominis, Phys. Lett. A {\bf 372}, 4877 (2008).

\bibitem{romalis_prl}
J. C. Allred, R. N. Lyman, T. W. Kornack and M. V. Romalis, Phys. Rev. Lett. {\bf 89}, 130801 (2002).

\bibitem{nature_magn}
I. K. Kominis, J. C. Allred, T. W. Kornack and M. V. Romalis,
Nature {\bf 422}, 596 (2003).

\bibitem{carminati}
R. Carminati and J. J.  S\'aenz, Phys. Rev. Lett. {\bf 84}, 5156 (2000).

\bibitem{kubala}
B. Kubala and J. K\"onig, Phys. Rev. B {\bf 67}, 205303 (2003).

\bibitem{pascazio_review}
P. Facchi and S. Pascazio, J. Phys. A: Math. Theor. {\bf 41}, 493001 (2008).

\bibitem{shimizu}
K. Koshino and A. Shimizu, Phys. Rep. {\bf 412}, 191 (2005).

\bibitem{kurizki}
A. G. Kofman and G. Kurizki, Nature {\bf 405}, 546 (2000).

\bibitem{hab}
R. Haberkorn, Molec. Phys. {\bf 32}, 1491 (1976).

\bibitem{wiseman}
H. M. Wiseman, Quantum Semiclass. Opt. {\bf 8}, 205 (1996).

\bibitem{jortner}
J. Jortner, J. Am. Chem. Soc. {\bf 102}, 6676 (1980).

\bibitem{kobori}
Y. Kobori {\it et al.}, Proc. Natl. Acad. Sci. USA {\bf 102}, 10017 (2005).

\bibitem{braginsky}
V. B. Braginsky and F. Y. Khalili, {\it Quantum Measurement}, Cambridge University Press, Cambridge, 1995.

\bibitem{itano}
W. M. Itano, Phys. Rev. A {\bf 41}, 2295 (1990).

\bibitem{gurvitz}
S. A. Gurvitz, Phys. Rev. B {\bf 56}, 15215 (1997).

\bibitem{steck}
K. Jacobs and D. A. Steck, Contemp. Phys. {\bf 47}, 279 (2006).

\bibitem{anglos}
D. Anglos, {\it Photoinduced Intrapeptide Electron Transfer Involving Novel Donor and Acceptor Amino Acids: A Triplet State Approach}, Ph.D. Thesis, Cornell University, 1994.

\end{thebibliography}
\end{document}